\documentclass[preprint,aps,11pt]{revtex4}

\usepackage{mathptmx}
\usepackage{graphicx}
\usepackage{times}
\usepackage{amsmath, amsthm, amssymb}
\usepackage{url}
\usepackage{subfig}
\usepackage{hyperref}
\usepackage{algorithm}
\usepackage{algorithmic}

\usepackage{lscape}

\begin{document}

\title{Structure and stability of online chat networks built on emotion-carrying links}
\thanks{All data used in this study are fully anonymized.}%

\author{Vladimir Gligorijevi\'c$^1$, Marcin Skowron$^2$, Bosiljka Tadi\'c$^1$}
\affiliation{$^1$Department of theoretical physics, Institute Jo\v{z}ef Stefan, Jamova 39, 1000 Ljubljana, Slovenia; $^2$Austrian Research Institute for Artificial Intelligence, Freyung 6/6, 1010 Vienna, Austria}

\date{\today}

\begin{abstract}
High-resolution data of online chats are studied as a physical system in laboratory in order to quantify collective behavior of users. Our analysis reveals strong regularities characteristic to natural systems with additional features. In particular, we find self-organized dynamics with long-range correlations in user actions and  persistent associations among users that have the properties of a social network.  Furthermore, the evolution  of the graph and its architecture with  specific k-core structure are shown to be related with the type and  the emotion arousal of exchanged messages. Partitioning of the graph by deletion of the links which carry  high arousal messages exhibits critical fluctuations at the percolation threshold.
\end{abstract}

\maketitle

\section{Introduction}
Quantitative analysis of human collective dynamics has recently become available based on the high-resolution empirical data from online communication systems, which can be studied as complex dynamical systems in physics laboratory. 
 The self-organized dynamics, common to the online interactions, is crucial for the emergence of collective behaviors of users \cite{castellano2009,kleinberg2008,mitrovic2010c}. The kind of social structures  built in the dynamics, however, may depend on the technology features of the communication platform and other details such as  visibility of user-to-user messages. Such details may affect actions of individual users and thus influence the course of events. 
For instance, the dynamics of message exchange along ``friendship'' links in the social network \texttt{MySpace} has  been shown \cite{we-MySpace11} to yield structures much different from the conventional social networks \cite{newman2002,boccaletti2006,facebook2011}. Apart from the online social networks, the  online games  \cite{szell2010,ST12}, blogs, diggs, forums \cite{mitrovic2011,mitrovic2010c,warsaw2011} etc, can also lead to recognizable user associations. In the online games \cite{szell2010,ST12}, for example, the  users can mark friend/enemy  relationship or undertake collective actions towards other users. On the other hand, indirect interactions on blogs provide hidden mechanisms in which the subjects of the posts and  negative emotion (critique) dominate, leading to user communities centered around certain popular posts \cite{mitrovic2011,mitrovic2010c}.
Furthermore, frequency of interactions,  subject of the communication, and amount of emotion conveyed in these messages may become critical for collective dynamics and for building  specific associations among users over time. These aspects of the online communication dynamics represent  a major challenge for quantitative analysis and theoretical modeling. Ideal empirical systems where this can be studied are the IRC (Internet-Relay-Chat) channels, where no \textit{a priori} relationship among users exists.

In this paper we combine  statistical physics with computer science methods of text analysis  to perform a quantitative analysis of human collective behavior in online chats. 
A large dataset from IRC \texttt{ Ubuntu channel} is analysed considering user-to-user communications with full text of messages  and assessing emotion contents in the text. In addition to humans, the data contains a Web robot (bot), which serves predefined text messages upon users' request. 
By mapping the data onto directed weighted network, and analyzing its topology in connection with temporal patterns of user actions, we find that a new type of techno-social structure emerges, which persists for  a long time. 
We further analyze its architecture by testing the validity of the ``social'' hypothesis and study the percolation transition that is connected with the amount of emotional arousal on the links. Both the graph architecture and its social ties suggest that a specific type of online social network is assembled based on the emotion-carrying communications.

\textit{Data structure.} We consider data from the publicly accessible IRC channel related to the development of the \texttt{Ubuntu} operating system \cite{http}. The data contain both humans and a bot. Users, identified by IDs, typically exchange short text messages in seeking information on software or services, joining group discussions, or conducting open-domain chats. 
The data considered in this study are for one year period (year 2009), and contain texts of messages and time resolution of one minute. After retrieval of all the logs for a given time window, anonymization by substituting user IDs by random number references, and removing the spam, the utterances are annotated with appropriate sentiment detecting tools. In particular,
in this study we use the lexicon-based annotation, by which emotion carrying words in a message are detected \cite{Bradley1999}. Such words carry emotional arousal---degree of reactivity, and valence---pleasure or displeasure,  listed  in the range $(1-9)$ in the emotional dictionary  \cite{Bradley1999}, from which  then the average arousal and valence are computed for the whole message.
Moreover, the content of each message is determined according to one (or more) dialog act classes. In total, twelve  such classes  can be determined  \cite{skowron_paltoglou2011affect}. For instance, three classes that  we consider here \textit{(yes-no question, why-question,  statement)}, in common ``question''-type messages, appear to be  characteristic to a certain group of users.

\section{Self-Organized Dynamics of Chats and Network Evolution}
\begin{figure}[h]
\begin{tabular}{cc}
\resizebox{18.8pc}{!}{\includegraphics{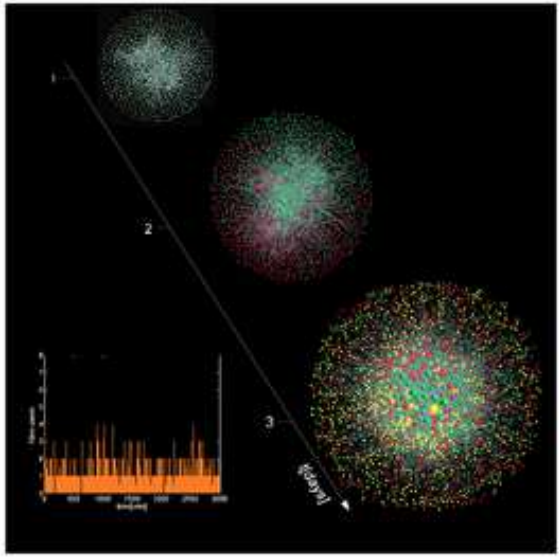}}
\end{tabular}
\caption{Daily growth of the directed network of chats by  addition of new users (time series in the inset) and by addition  of links and increase of the links intensity. }
\label{fig-net-evolution}
\end{figure}

The data are filtered to identify unique user IDs (the ID labeled by ``35177'' belonging to the bot is treated equally). The subset where  user-to-user communications are clearly identified is selected and mapped onto a directed network. 
The network nodes $i=1,2\cdots N$ represent users and the directed link $i\to j$ indicates that at least one message from user $i$ to user $j$ occurred within the considered time window. Multiple messages along the link increase its width, while the message emotion and message category are considered  as \textit{properties} of that link. Starting from the beginning of the dataset, the network evolves in time by addition of new users and new links and increasing the widths of the existing links.  Fig.\ \ref{fig-net-evolution} shows how the network grows within first three days.  After one year (data limit) the network consists of $N=$ 85185 users with  the link density $\rho \equiv L/N(N-1)=1.244\times 10^{-4}$ and link reciprocity $r\equiv L^{\leftrightarrow}/L =0.284$.

\begin{figure}[h]
\begin{tabular}{cc}
\resizebox{18.8pc}{!}{\includegraphics{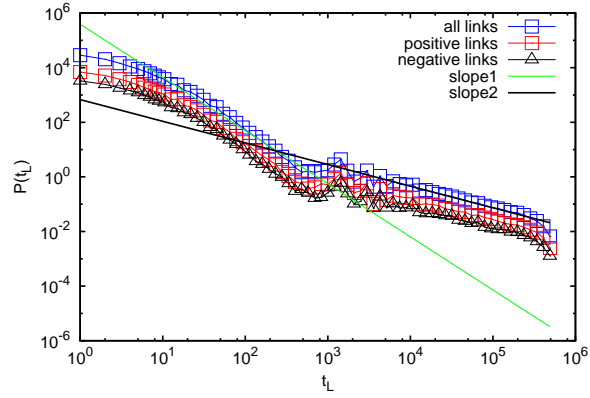}}\\
\end{tabular}
\caption{Distribution of the lifetime of links with positive and  negative overall emotion messages, and the links with all types of messages.
}
\label{fig-lifetime-links}
\end{figure}

In the temporal patterns of user actions we noticed that some users appear for a short time and disappear from the channel, while others are constantly or occasionally active within a long time period. We examine the \textit{lifetime of links} that the users establish between each other. The distribution of the lifetimes of all links in the dataset is shown in Fig.\ \ref{fig-lifetime-links}.
The three curves are, respectively, for all kinds of links, and for the links that carry an overall positive and negative emotion valence. Apart from the frequency, the patterns are similar: For lifetimes shorter than one day, the distribution decays faster than for the lifetimes longer than one day. The power-law with a small exponent of the distribution in the latter region suggests that the links that survive the first day after the appearance are likely to persist for a longer time (till data limit). The network that  contains only such persistent links, here termed the \textit{persistent network}, or \texttt{UbuNetP}, is of our interest and can be considered as a potential social structure. 
Thus, we remove the links which do not survive over the first day after their appearance. Note that in this way a number of  users are also removed. 
(The user survival distribution follows a similar pattern, not shown).
In the following we explore the emergent network with the persistent links and study how the above mentioned link properties, in particular the message types and their emotion contents,  are related with the network architecture. 
\begin{figure}[h]
\begin{tabular}{cc}
\resizebox{18.8pc}{!}{\includegraphics{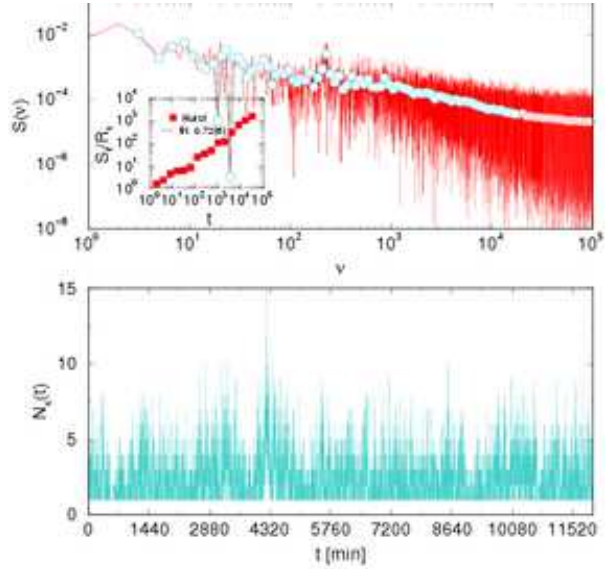}}
\end{tabular}
\caption{ (bottom) Time series of the number of messages  and (top) its power spectrum. Inset:  Largest fluctuations of the integrated time series scaled by its standard dispersion plotted against time interval.  }
\label{fig-timeseries}
\end{figure}

These network connections emerge through a self-organized dynamics, in which users play different functions and over time they ``settle'' in the evolving environment. 
The cooperative behavior is quantified by analysis of correlated fluctuations    in the time series of user actions. Specifically, we construct the time series of the number of messages per one minute  time bin. In Fig.\ \ref{fig-timeseries} an example of the time series is shown in the bottom panel. For better view,  only a small part  corresponding to eight days is shown, while the length of the considered time series is $T=262144$ bins. 
The power spectrum of the time series is given in the top panel in Fig.\ \ref{fig-timeseries}. The inset shows the fluctuations $R_t/\sigma_t\sim t^H$, scaled by the standard deviation $\sigma_t$, of the integrated time series $Y_t  =\sum_{\tau=1}^t[x(\tau)-<x(\tau)>]$ as a function of varying time window $t =1,2, \cdots T$. The slope determines the Hurst exponent $H$. Apart from the peak corresponding to daily periodicity, the spectrum shows long-range correlations of the Gaussian noise type with $S(\nu)\sim 1/\nu^\phi$ in a wide range of frequency index $\nu$, which is indicated by the straight line. The correlation range, considered in the time domain, agrees well with the persistence time window of the fluctuation, i.e., $t\in [5,32000]$. Note also that the values of the exponents $\phi=0.48\pm 0.03$ and $H=0.72\pm 0.06$ satisfy (within error bars) the  scaling relation $\phi=2H-1$ expected for the colored noise signals. Furthermore, the value of the Hurst exponent $H>1/2$ suggests that a \textit{persistent} type of fluctuations occurs in the system as a whole.  T
Note that the Hurst exponent in this range is also found for the time series of messages carrying emotional content along a particular link or by a particular user \cite{we-Chats12,garas2011}.

\section{Content-Based Structure and Resilience of the Social Graph}

The network with persistent links is further examined as a ``social'' structure. We recall that  ``weak-ties'' hypothesis has recently been  confirmed in the data of mobile phone, online games, and \texttt{MySpace} dialogs networks \cite{onela2007,szell2010,ST12,we-MySpace11}. 
In analogy to the social networks, the test is performed by computing the overlap $O$---the number of common neighbors  of two nodes which share a link, as a function of the width $W$ and of the betweenness centrality $B$ of that link.
 In particular, the overlap increases with the intensity of communications $W$  between the nodes as  $O(W)\sim W^{\eta_1}$, while it decreases with the centrality of the links as $O(B)\sim B^{-\eta_2}$. Both laws hold in the above mentioned online systems, all of which have a typical  \textit{community structure}. The corresponding exponents are close to the 
 universal values $\eta_1=1/3$ and $\eta_2=1/2$, which are recently conjectured in \cite{ST12}. 

\begin{figure}[h]
\begin{tabular}{cc}
\resizebox{18.8pc}{!}{\includegraphics{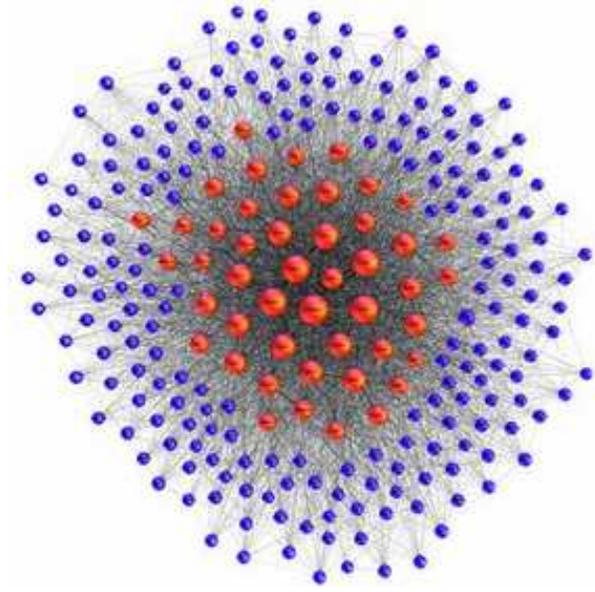}}\\
\resizebox{18.8pc}{!}{\includegraphics{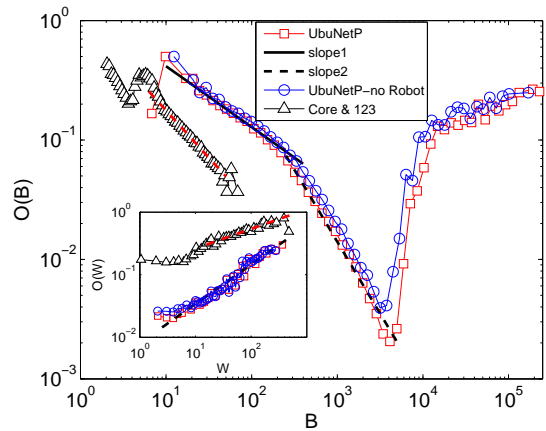}}\\
\end{tabular}
\caption{ Top: Network with central core nodes (red) and  users whose activity is dominated by question-type messages (blue).
Bottom: The overlap $O$ of the adjacent nodes plotted against the  betweenness centrality $B$, main figure, and  the weight $W$ of their link, inset, for the persistent network \texttt{UbuNetP} and two its subgraphs, described in the text.}
\label{fig-Ubu-overlaps}
\end{figure}
In the emergent networks of chats, the situation is somewhat different in that no classical community structure is found. Moreover, the network shows hierarchical organization of links with a large ``core'', but the links between nodes at different hierarchy levels also occur. Thus the network remains connected even if the  central core is removed \cite{comment}. 
Consequently, we find that the overlap has a different dependence of betweenness centrality,  as shown in Fig.\ \ref{fig-Ubu-overlaps}. The inset shows the overlap as function of $W$. The three curves are for our persistent network \texttt{UbuNetP}, and for two subnetworks described below. One subnetwork is obtained by removing the bot and all its links. The other is a reduced network shown in the top Fig.\ \ref{fig-Ubu-overlaps}, whose  structure reflects the type of messages that the users exchange: the outer layer (blue nodes) represents the users who send question-type of messages twice more often than the average frequency on the channel, and no other message types. They are attached to the central core (red nodes), which consists of  the bot and the users who frequently use all types of messages. Qualitatively,  the ``social ties'' behaviors are present, but the corresponding exponents are twice larger, $\eta_1\approx 0.66$ and $\eta_2\approx 1$,  except for a small parts of the curves (marked by the red dashed line), which agree with the other online social systems. The origin of such behavior as well as the sudden increase of the overlap at high $B$ values needs theoretical modeling and analysis of more similar systems, which is left out of the present work. It is interesting  to note that  qualitatively same  behavior is found in the network from which the bot is removed.

\begin{figure}[h]
\begin{tabular}{ccc}
\resizebox{24.8pc}{!}{\includegraphics{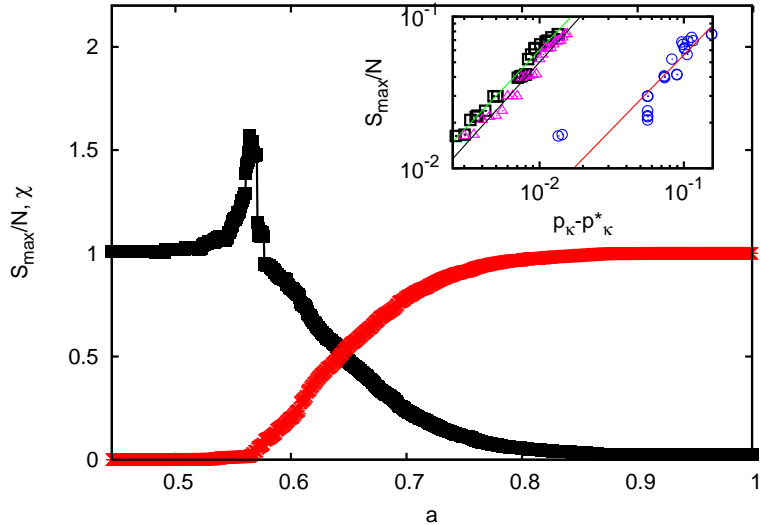}}\\
\end{tabular}
\caption{Relative size of the giant cluster $S_{max}/N$ and the susceptibility $\chi$ plotted against rescaled threshold arousal of the links $a$. Inset: $S_{max}/N$ against distance from the percolation point $p_\kappa -p_{\kappa}^\star$, $\kappa \in(a, p, q)$.}
\label{fig-Ubu-arousal-percolation}
\end{figure}

The chat network has a structure with a power-law distribution for in- and out-degree with a similar exponent $\tau_q=2.0 \pm 0.02$ and disassortative mixing. A detailed analysis  is left for a separate study \cite{we-Chats12}. Hereafter we focus on the resilience of the persistent chats network by analyzing the \textit{percolation transition} that occurs in relation to the arousal of the messages carried along the directed links.  The rationale is that the links persist over a long time due to certain attribute (or ``excitement'', properly quantified by the emotional arousal) of the messages exchanged along them. It is worth noting that the  weights and the averaged arousal of the links obey  a  power-law and a skewed Gaussian distribution, respectively  \cite{we-Chats12}. 
Starting from the whole persistent network (with $N =6168$ users and $L=33838$  links) we  cut the directed links  whose average  arousal   exceeds a threshold value, i.e.,  $\overline{a_{ij}}/\overline{a}_{max}\geq a$. For convenience it is  normalized  by $\overline{a}_{max}$, the largest average arousal found on the network. 
Note that, depending on the actual value of the threshold $a$, the number of links that satisfy the condition as well as their position on the network can vary considerably. Together with the network topology, directedness, and reciprocity of the links, this fact of the content-based topology determines the features of the percolation transition \cite{boguna2005}.
For each threshold $a$ the largest cluster (as weakly connected component) is found and its size $S_{max}/N$  is plotted against the threshold $a$ in Fig.\ \ref{fig-Ubu-arousal-percolation}. 
Also shown is the susceptibility $\chi$, which is defined as $\chi = \sum_{s< S_{max}}s^2n_s$. It determines the fluctuations of the cluster size in response to cutting the links with a given arousal. The distribution of cluster size satisfies the power law $n_s \sim s^{-2.6\pm 0.08}$.
In agreement with the percolation theory, $\chi$  exhibits a peak at the transition point $a^\star=0.561$, where the $S_{max}/N$ tends to vanish. 

What kind of geometry and/or complexity change \cite{dorogovtsev2008,schwartz2002,boguna2005,neelima2011} in the network structure occurs at the transition point  $a^\star$? 
We examine  the fraction of retained links $p$, and the 
average degree $q$ of the giant cluster, for the range of values of $a\gtrsim a^\star$ above the transition point. In the inset of Fig.\ \ref{fig-Ubu-arousal-percolation} we show the scaling region $S_{max}/N \sim (p_\kappa -p_\kappa^\star)^\beta$, where the index indicates $\kappa \equiv (a,p,q)$.   The critical value $p^\star=0.015$ and  $q^\star =1.944$. The exponent $\beta$ is close to one, i.e., $\beta=1.06\pm 0.04$, $1.00\pm 0.05$,  $1.0\pm 0.1$ for  $\kappa = a,p,q$, respectively.

\section{Conclusion}
 Applying physics methods we have studied the self-organized dynamics of user-to-user interactions in the online chats, and have shown that both the  type of the messages and their emotional arousal play an  essential role in building a persistent techno-social structure. 
The social graph is  hierarchically  organized having a central core with some very active users and the Web bot. 
In contrast to more familiar user grouping into communities, this kind of spontaneous organization marks a new class of structures in the zoo of emergent online communication networks. 
The observed percolation transition on this graph is smooth with the enhanced  fluctuations  in response to a targeted deletion of links  which carry  the emotional arousal over a threshold value.  
Our analysis reveals that the contents of the exchanged messages and the role that users assume in the dynamics need to be considered as an integral parts of the topology in the online social graphs.
In general, the idea to consider features beyond conventional interactions, that we pursued here for the analysis of human online interactions, can also be useful for studying physical systems which are assembled from complex molecules or other objects in laboratory. 

{\bf Authors' contributions:} BT designed research, analyzed data and wrote the paper; VG contributed software and performed analysis; MS collected data and performed annotation of emotional contents.

{\bf Acknowledgments:}  {
We thank the support from the program P1-0044 by the  Research agency of the Republic of Slovenia and from the European Community's program  
FP7-ICT-2008-3 under grant n$^o$ 231323.  }


\begin{thebibliography}{10}
\bibitem{comment} Considered as an undirected graph, the size of $k$-core is found to decreases as power $k^{-1.47\pm0.03}$. The strongest core is  $k=40$, which also contains the bot.
\end{thebibliography}

\end{document}